\documentclass[referee]{chjaa}          

\usepackage{graphicx,times}             
\input{epsf.sty}                        
\input{psfig.sty}                       

\begin{document}

   \title{A Reinvestigation of the Physical Properties of Pismis 3 based on
2MASS Photometry}

   \volnopage{Vol.0 (200x) No.0, 000--000}      
   \setcounter{page}{1}           

   \author{Tadross, A. L.
      \inst{}\mailto{}
        }

   \institute{National Research Institute of Astronomy and
Geophysics, 11421 - Helwan, Cairo, Egypt\\
             \email{altadross@nriag.sci.eg}
       }

   \date{Received~~2007 month day; accepted~~2007~~month day}
Chin. J. Astron. Astrophys. Vol. 8 (2008), No. 3, 362\\
Received 2007 May 31; accepted 2008 April 11
   \abstract{
As a continuation of a series of work, we aim to refine and re-determine the physical
parameters of previously rarely or un-studied open star clusters with good quality CMDs
using Near-IR JHK photometry. Here we present a morphological analysis of the 2MASS
database (the digital Two Micron All Sky Survey) for the open cluster Pismis 3. Some of
the physical parameters are estimated for the first time, and some others, re-determined.
   }

   \authorrunning{Tadross, A. L.}            
   \titlerunning{A Reinvestigation of the Physical Properties of Pismis 3}  

   \maketitle

%
%
\section{Introduction}           

A deep photometric and astrometric analysis in the open star
cluster Pismis 3 has been presented here using {\it
2MASS}\footnote{\it http://www.ipac.caltech.edu/2MASS} database.
The {\it 2MASS} Surveys has proven to be a powerful tool in the
analysis of the structure and stellar content of open clusters
(cf. Bonatto \& Bica 2003, Bica et al. 2003). It is uniformly
scanning the entire sky in three near-IR bands $J$(1.25 $\mu$m),
$H$(1.65 $\mu$m) and $K$$_{s}$(2.17 $\mu$m) with two highly
automated 1.3-m telescopes equipped with a three channel camera,
each channel consisting of a 256$\times$256 array of HgCdTe
detectors. The photometric uncertainty of the data is less than
0.155 mag with $K_s \sim$ 16.5 mag photometric completeness.
Further details can be found at the web site of {\it 2MASS}.
\\
Pismis 3 (C0829-3830, OCL 731) is situated in the southern Milky
Way at 2000.0 coordinates $\alpha=08^{h} \ 31^{m} \ 22^{s},
\delta=-38^{\circ} \ 39^{'} \ 00^{''}; \ \ell= 257.865^{\circ}, b=
+0.502^{\circ}$. Carraro \& Ortolani (1994), hereafter CO94,
obtained {\it CCD BV} photometry for Pismis 3 and its nearby
field. Their analysis suggests that it is of intermediate age
(about 2 Gyr) and metal poor (Z = 0.008) cluster. They derived a
color excess E(B-V) = 1.35, and an apparent distance modulus (m-M)
= 14.70 mag (about 1.5 Kpc distant from the Sun). In our series,
the most fundamental parameters have been estimated, i.e. age,
reddening, distances (from the sun; the galactic plane; the
galactic center), diameters (cluster's border; core radius; tidal
radius), luminosity function, mass function, total mass,
relaxation time, and mass segregation. Relevant examples are NGC
1883; NGC 2059; NGC 7086 (Tadross 2005), and NGC 7296 (Tadross
2006). This paper is organized as follows: Sect. 2, data
extraction; Sect. 3, cluster center \& radii; Sect. 4, {\it CMD}
analysis (membership richness - reddening - distances - age, and
metallicity); Sect. 5, luminosity function; Sect. 6, mass
functions \& total mass; Sect. 7, mass segregation \& dynamical
state; and finally the conclusions have been summarized and listed
with a comparing table in Sect. 8. Fig. 1 represents the blue
image of Pismis 3 as taken from Digitized Sky Surveys {\it
(DSS)}\footnote {\it http://cadcwww.dao.nrc.ca/cadcbin/getdss}. It
takes the name from the astronomer who in the late fifties (Pismis
1959) who compiled a catalogue of 2 globular and 24 new open star
clusters in the Galactic Plane between $\ell = 225^{\circ}$ and
$\ell = 353^{\circ}$ (cf. CO94).

\begin{figure}
   \vspace{2mm}
   \begin{center}
   \hspace{3mm}\psfig{figure=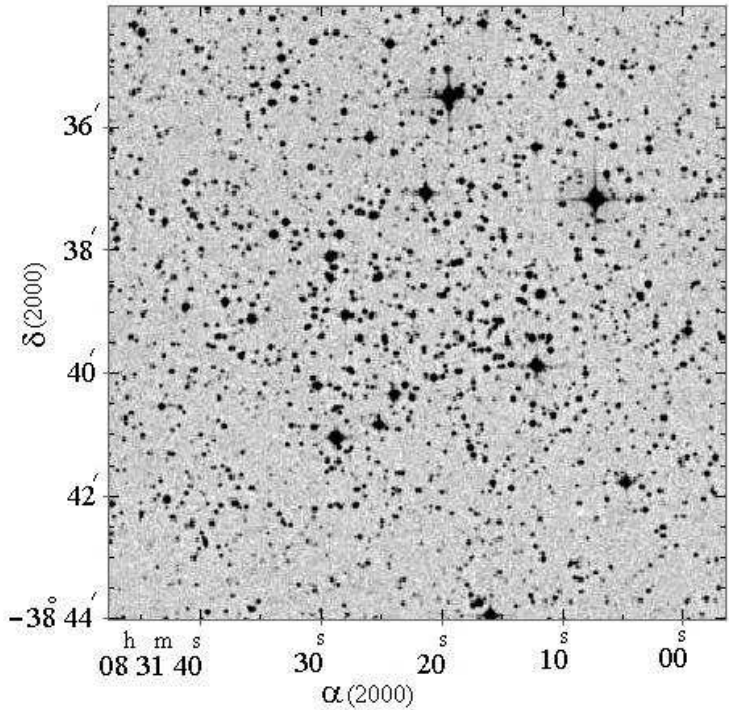,width=100mm}
   \caption{The blue image of Pismis 3
as taken from Digitized Sky Surveys {\it (DSS)}. North is up, east
on the left.}
   \label{Fig:}
   \end{center}
\end{figure}

\section{Data extraction}

Data extraction have been performed using the known tool of
VizieR\footnote{\it
http://vizier.u-strasbg.fr/viz-bin/VizieR?-source=2MASS}. The
number of stars in the direction of Pismis 3 within a preliminary
radius of 10 arcmin is found to be 4390 stars. In order to
maximize the statistical significance and representativeness of
background star counts, an external area (the same area as the
cluster) has been used as offset field sample. This external
sample lies at 1 degree away from the cluster's center.
\\ \\
Before counting stars for estimating the cluster's properties with
\emph{JHK} \emph{2MASS} photometry, we applied a cutoff of
photometric completeness ($J<16.5$) to both cluster and offset
field to avoid over-sampling, i.e. to avoid spatial variations in
the number of faint stars which are numerous, affected by large
errors, and may include spurious detections (Bonatto et al. 2004).
Also, in this respect, for more accuracy, we restricted to stars
with observational uncertainties $\epsilon_{~J,~H,~K}<0.2$\,mag.

\begin{figure}
   \vspace{2mm}
   \begin{center}
   \hspace{3mm}\psfig{figure=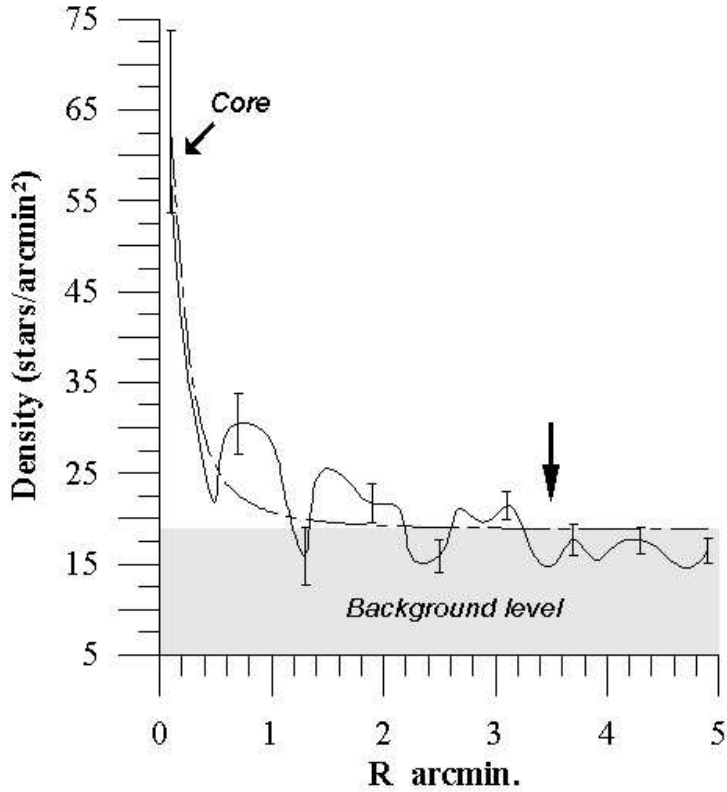,width=100mm}
   \caption{Radial distribution of the
surface density of Pismis 3 (solid curve). The vertical short bars
represent the Poisson errors. The dashed line represents the
fitting of King (1962). The upper arrow refers to the core region,
and the lower one marks the apparent minimum radius of the
cluster. The dark region represent the mean level of the offset
field density, which taken at $\sim$ 18 stars per arcmin$^{2}$.}
   \label{Fig:}
   \end{center}
\end{figure}

\section{The cluster center and radii}

The cluster center is define as the location of maximum stellar
density of the cluster's area. The cluster center is found by
fitting a Gaussian to the profiles of star counts in right
ascension ($\alpha$) and declination ($\delta$), see Tadross 2004,
2005 and 2006. The estimated center is found to lie at $\alpha$ =
127.84089 $\pm$ 0.003 and ~~~ $\delta$ = -38.64478 $\pm$ 0.002
degrees, which is found to differ from WEBDA\footnote{\it
http://obswww.unige.ch/webda/navigation.html} by 0.2 sec in right
ascension and 18.8 arcsec in declination.\\

To determine the cluster's minimum radius, core radius and tidal
radius, the radial surface density of the stars $\rho(r)$ should
be achieved firstly. The tidal radius determination is made
possible by the spatial coverage and uniformity of {\it 2MASS}
photometry, which allows one to obtain reliable data on the
projected distribution of stars for large extensions around
clusters (Bonatto et al. 2005). In this context, the background
contribution level corresponds to the average number of stars
included in the offset field sample is found to be $\sim$ 18 stars
per arcmin$^{2}$. Applying the empirical profile of King (1962),
the cluster's minimum apparent radius is taken to be 3.5 arcmin,
as shown in Fig. 2. Knowing the cluster distance from the sun in
parsecs (\S~4), the cluster and core radii are found to be 2.2 and
0.2 pc respectively. Applying the equation of Jeffries et al.
(2001), the tidal radius ($R_t$) of Pismis 3 is found to be $\sim$
12 pc. Consequently, the distances of the cluster from the
galactic plane, Z, and its projected distances from the Sun,
$X_{\odot}$, $Y_{\odot}$, are found to be 18.0 pc; --2.0, and 0.44
\,kpc respectively. The distance from the galactic center, $R_g$,
is found to be 8.7, or 8.0 kpc using the galactocentric distance
of the sun $R_o$ = 8.0, or~ 7.2 kpc according to Reid (1993), or
Bica et al. (2006) respectively. It is found that $R_g$ is
consenting with; but Z is larger than what obtained by Salaris et
al. (2004).

\begin{figure}
   \vspace{2mm}
   \begin{center}
   \hspace{3mm}\psfig{figure=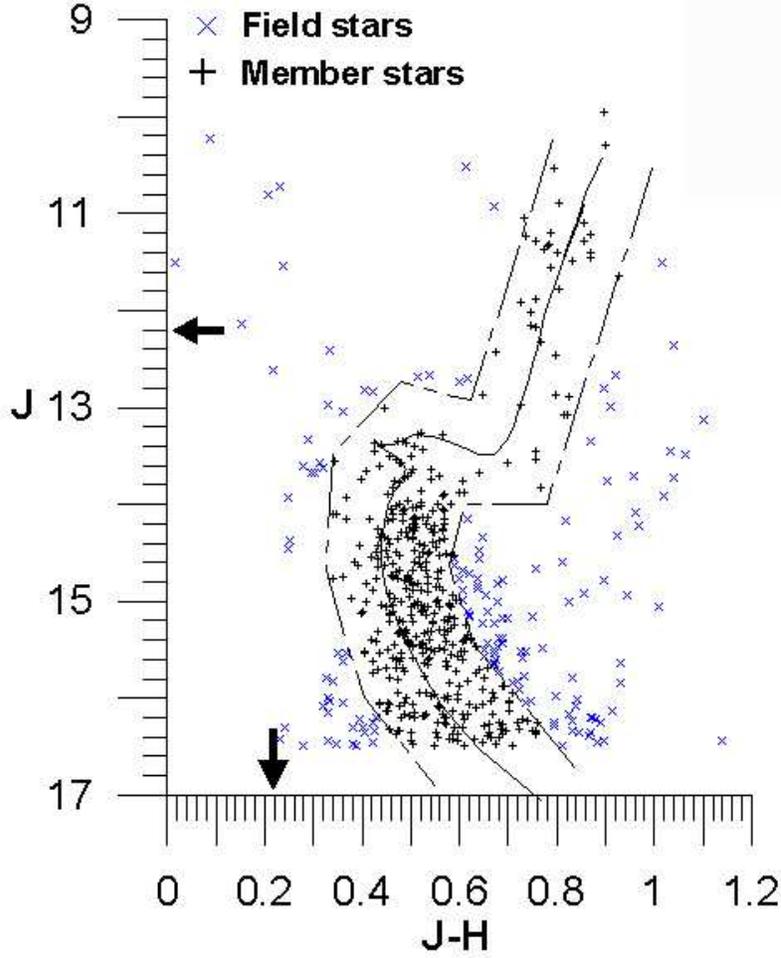,width=105mm}
   \caption{Padova solar isochrone
with log age = 9.35 (2.24 Gyr) is fitted to the $J\sim(J-H)$ {\it
CMD} of Pismis 3. Dashed curves represent the color and magnitude
filters used in reducing the field contamination of the cluster.
The horizontal and vertical arrows refer to the values of distance
modulus and color excess on the vertical and horizontal axes
respectively.}
   \label{Fig:}
   \end{center}
\end{figure}

\section{Color-Magnitude Diagram analysis}

Because of the low galactic latitude of Pismis 3, the background
field of the cluster is found to be crowded ($\approx$ 18 stars
per arcmin$^{2}$), and the observed {\it CMD} is contaminated.
Fig. 3 represents the {\it CMD} of Pismis 3, showing the magnitude
completeness limit and the color filter for the stars within the
apparent cluster radius, whereas 450 stars are classified as
cluster members. The membership criteria here is adopted for the
location of the stars in the {\it CMD}, which must be close to the
cluster main sequence (the stars lie between the two dashed curves
in Fig. 3, which have "+" signs), the maximum departure accepted
here is about 0.15 mag. On this base, the fundamental
photometrical parameters of the cluster (reddening, distance
modulus, age, and metal content) can be determined simultaneously,
by fitting one of Padova isochrones to the {\it CMD} of the
cluster.

In this respect, several fittings have been applied on the
$J\sim(J-H)$ of Pismis 3 using Bonatto et al. (2004) isochrone of
solar metallicity with different ages. $\rm R_V=3.2$, $\rm
A_J=0.276\times A_V$, and $E(J-H)=0.33 \times E(B-V)$ have been
used for reddening and absorption transformations, according to
Dutra, Santiago \& Bica (2002) and references therein. The overall
shape of the {\it CMD} is found to be well reproduced with
isochrones of 2.24 Gyr in age. The apparent distance modulus is
found to be 12.20 $\pm$ 0.10 mag, accordingly the intrinsic one,
$(m-M)_{o}$, is found to be 11.60 $\pm$ 0.10 mag, corresponding to
a distance of 2090 $\pm$ 95 pc. On the other hand, the color
excess, $E(J-H)$, is found to be 0.22 mag, which turns out to be
$E(B-V)$ = 0.67 mag. It is found that [Fe/H] is consenting with;
but the age is smaller than what obtained by Salaris et al.
(2004). It is worthily to mention that the noticed differences of
the main parameters for Pismis 3 between the present work and CO94
is mainly due to the difference of the metal content of the used
isochrone.

\begin{figure}
   \begin{center}
    \hspace{3mm}\psfig{figure=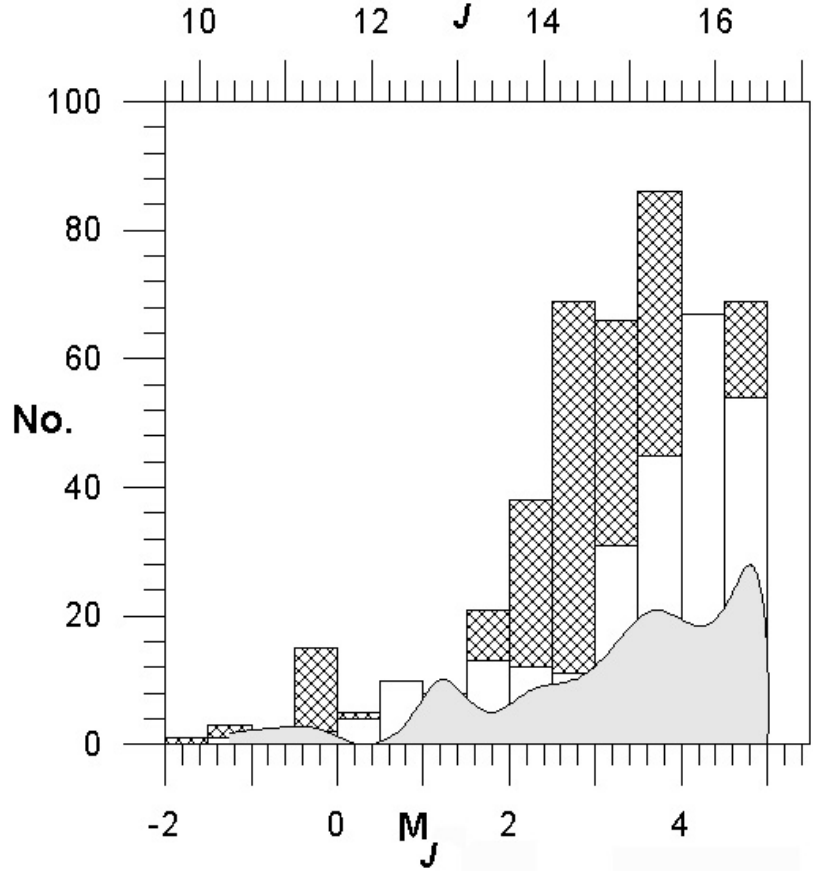,width=115mm}
     \caption{Spatial distribution of
luminosity function for Pismis 3 in terms of the absolute
magnitude $M_{J}$. The color and magnitude filters cutoffs have
been applied to the cluster (dashed area) and the offset field
(white area). The dark curved area represents the background
subtracted {\it LF}. The scale of observed J magnitude appears
along the upper axis. }
   \end{center}
\end{figure}

\section{Luminosity function}

The observed stars have been counted in terms of the absolute
magnitude $M_{J}$ after applying the distance modulus derived
above. The color and magnitude filters cutoffs have been applied
to the cluster and offset field stars. The magnitude bin interval
are taken to be $\Delta M_{J}=0.50$\,mag. In Fig. 4, the {\it LF}
constructed as the difference in the number of stars in a given
magnitude bin between the cluster's stars (dashed area) and the
offset field ones (white area). Dotted area represents the
background subtracted {\it LF}. The scale of observed J magnitude
appears along the upper axis of Fig. 4. From the {\it LF} of
Pismis 3, we can infer that more massive stars are more centrally
concentrated whereas the beak value lies at fainter magnitude bin
(Montgomery et al. 1993). This peak corresponds to J $\approx$
15.3 mag, i.e. M$_{J}$ $\approx$ 3.7 mag.

\section{Mass function and total mass}

Given the luminosity function, the mass function and then the
total mass of the cluster can be derived. To derive the {\it MF}
from {\it LF}, the theoretical evolutionary track of Bonatto et
al. (2004) with solar metal abundance (Z=0.019) and age of 2.24
Gyr is used. In this sense, a polynomial equation of fourth
degrees has been used for the cluster members in the range of
-1.75 $\leq M_{J} \leq$ 4.75 as following:

\begin{center}
$\mathcal{M}$/$\mathcal{M}_{\odot}$=~3.13--~0.66~M$_{J}$--~0.10~M$_{J}^{2}$+
0.051~M$_{J}^{3}$--~0.005~M$_{J}^{4}$
\\
\end{center}
Step-plot has been constructed for the cluster stellar masses
showing the number of stars at 0.5 intervals between 0.65 $\sim$
3.65 $\mathcal{M}_{\odot}$, as shown in Fig. 5. Using a
least-square fit, the slope of {\it IMF}~ is found to be $\Gamma$
= -2.37$\pm$ 0.25, which is about in agreement with the value of
Salpeter (1955). In this respect, the total mass of the cluster
has been estimated by summing up the stars in each bin weighted by
the mean mass of that bin. It yields a minimum cluster mass of
$\sim 560\,\mathcal{M}_{\odot}$.

\begin{figure}
   \begin{center}
   \hspace{3mm}\psfig{figure=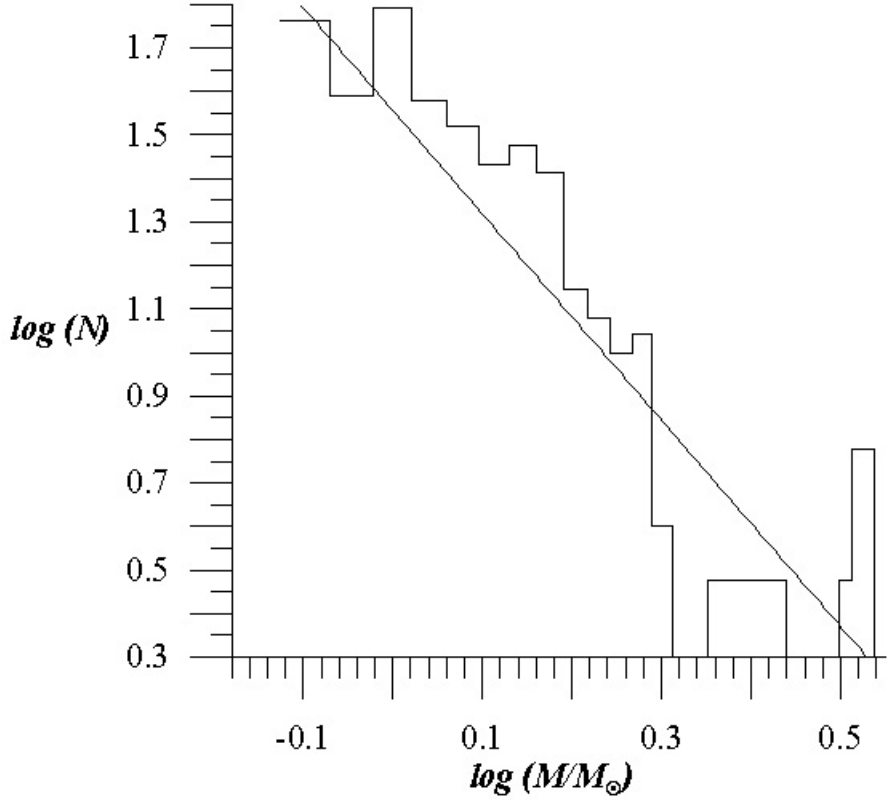,width=120mm}
   \caption{The mass function of
Pismis 3. The slope of the initial mass function {\it IMF}~ is
found to be $\Gamma = -2.37 \pm 0.25$; with correlation
coefficient of 0.90.}
   \end{center}
\end{figure}

It is noted that, unresolved binaries and low mass stars are
problems for this technique. In this respect, Van Albada \& Blaauw
(1967) assumed that 60\% of early type stars are double systems,
whereas Jaschek \& Gomez (1970) claimed that approximately 50\% of
the main sequence stars might be hidden (cf. Bernard \& Sanders
1977). According these assumptions, the total mass of the cluster
Pismis 3 can be reached to $\sim$ 800 $\mathcal{M}_{\odot}$.

\begin{figure}
   \begin{center}
   \hspace{3mm}\psfig{figure=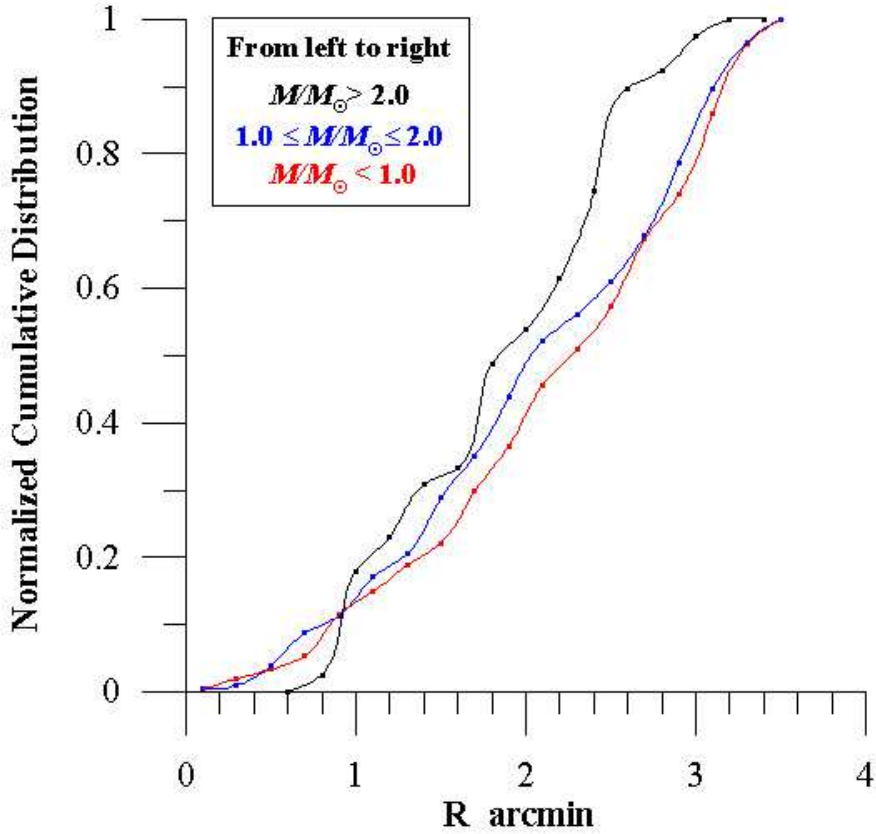,width=120mm}
   \caption{Mass segregation in Pismis
3. Moving from left to right, the curves represent the mass ranges
$\mathcal{M}/\mathcal{M}_{\odot}>2.0;(M_{J}:-0.1\sim1.5),1.0\leq
\mathcal{M}/\mathcal{M}_{\odot}\leq 2.0;(M_{J}:1.6\sim3.6)$, and $
\mathcal{M}/\mathcal{M}_{\odot}<1.0;(M_{J}:3.6\sim4.9).$ This
indicates that the bright massive stars accumulate much more
quickly with radius than the fainter low mass stars do.}
   \end{center}
\end{figure}

\section{Mass segregation and dynamical state}

For a dynamically relaxed cluster, the higher mass stars are
expected to be settled toward the cluster center, while the
fainter, lower mass stars are residing in the outer regions of the
cluster, Mathieu (1984). The existence of mass segregation is due
to the dynamical evolution or/and imprint of star formation
process. At the time of formation, the cluster may have a uniform
spatial stellar mass distribution, and because of the dynamical
relaxation, low mass stars may possess the largest random
velocities, trying to occupy a larger volume than the high mass
stars do (cf. Mathieu \& Latham 1986, McNamara \& Sekiguchi 1986,
Mathieu 1985).
\\ \\
To display mass segregation in Pismis 3, star counts are performed
on the main sequence as a function of distances from the cluster
center and masses. The results are given in Fig. 6. The individual
curves moving from left to right are for mass ranges
$\mathcal{M}/\mathcal{M}_{\odot}> 2.0, \, 1.0\leq
\mathcal{M}/\mathcal{M}_{\odot}\leq 2.0$, \,and
$\mathcal{M}/\mathcal{M}_{\odot}<1.0$. It suggests that the
brighter high mass stars concentrate towards the cluster center
and accumulate much more quickly than the fainter low mass stars
do. On the other hand, we are interested if the cluster reached
the dynamical relaxation or not. Applying the dynamical
relaxation' equation (cf. Tadross 2005 \& 2006), it is found to be
8.6 Myr, which implies that the cluster age is $\sim$ 260 times
its relaxation one. Thus we can conclude that Pismis 3 is
dynamically relaxed and the evolution is one of the possible cause
of mass segregation.

\section{Conclusions}

According our analysis for refining and
determining the fundamental parameters of Pismis 3 using {\it
2MASS} photometry, the present results are summarized and compared
with the previous one (CO94) in table 1.

\begin{acknowledgements}
This publication made use of the Two Micron All Sky Survey {\it
(2MASS)}, which is a joint project of the University of
Massachusetts and the Infrared Processing and Analysis Center
California Institute of Technology, funded by the National
Aeronautics and Space Administration and the National Science
Foundation. Catalogues from {\it $CDS/SIMBAD$} (Strasbourg), and
Digitized Sky Survey {\it DSS} images from the Space Telescope
Science Institute have been employed.
\end{acknowledgements}

\label{lastpage}

\begin{table}
\caption{Comparisons between the present study and CO94}
\begin{tabular}{lll}
\hline\noalign{\smallskip}Parameter&The present work& CO94
\\\hline\noalign{\smallskip}
Center&$\alpha$ = 08$^{h} 31^{m} 21.8^{s}$& 08$^{h} 29^{m} 6^{s}$\\
&$\delta$ = --38$^{\circ} 38^{'} 41.2^{''}$& --38$^{\circ} 30^{'} 0^{''}$\\
Age& 2.24 Gyr.& 2.0 Gyr.\\
Metal abundance& 0.019&0.008\\
$E(B-V)$& 0.67 mag.& 1.35 mag.\\
$R_{v}$& 3.2& 3.0\\
Distance Modulus& 12.20 $\pm$ 0.10 mag.& 14.70 mag.\\
Distance& 2090 $\pm$ 95 pc.& 1500 pc.\\
Radius& 3.5$^{'}$ (2.20 pc.)& 3.25$^{'}$ \\
Membership& 450 stars &--~--\\
$E(J-H)$& 0.22 mag. &--~--\\
$\rho_o$& 63$\pm$ 2 stars/arcmin$^{2}$& --~--\\
Core radius& 0.19$^{'}\pm 0.04$ (0.20 pc)& --~--\\
Tidal radius& 12 pc.& --~--\\
$R_g$& 8.7 $\sim$ 8.0 kpc. (see $\S$ 3)& --~--\\
Z& 18 pc.& --~--\\
X$_{\odot}$& --2.0 kpc.& --~--\\
Y$_{\odot}$& 0.44 kpc.& --~--\\
Luminosity fun.& Estimated& --~--\\
{\it IMF} slope& $\Gamma = -2.37 \pm 0.25$& --~--\\
Total mass& $\approx$ 560 $\mathcal{M}_{\odot}$ (minimum)& --~--\\
Relaxation time& 8.6 Myr& --~--\\
Mass segregation& Achieved& --~--\\
\hline
\end{tabular}
\end{table}

\end{document}